\documentclass{article}
\usepackage{amsmath}
\usepackage{amsfonts}
\usepackage{epsfig}

\newcommand{\cte}[1]{$^{\mbox{\scriptsize \cite{#1}}}$}
\newcommand{\hs}[1]{\hspace*{#1cm}}
\newcommand{\vs}[1]{\vspace*{#1cm}}
\newcommand{\half}{{\textstyle \frac{1}{2}}}
\newcommand{\bea}{\begin{eqnarray}}
\newcommand{\eea}{\end{eqnarray}}

\newcommand{\ca}{\cosh(\alpha)}
\newcommand{\seca}{\mbox{sech}(\alpha)}
\newcommand{\sa}{\sinh(\alpha)}
\newcommand{\ta}{\tanh(\alpha)}
\newcommand{\ct}{\cosh(a \tau_c)}
\newcommand{\st}{\sinh(a \tau_c)}
\newcommand{\sect}{\mbox{sech}(a \tau_c)}
\newcommand{\tant}{\tanh(a \tau_c)}

\begin{document}

\title{On Radar Time and the Twin `Paradox'}
\author{Carl E. Dolby\thanks{Email: c.dolby@mrao.cam.ac.uk} and Stephen F. Gull\thanks{Email: steve@mrao.cam.ac.uk} \\ 
{\em Astrophysics Group, Cavendish Laboratory, Madingley Road, Cambridge
CB3 0HE, U.K.} \\
{\bf Manuscript No: 12409 }}  \date{\today} \maketitle

\begin{abstract}

In this paper we apply the concept of radar time (popularised by 
Bondi in his work on k-calculus) to the well-known relativistic twin `paradox'. Radar time is 
used to define hypersurfaces of simultaneity for a class of travelling twins, from 
the `Immediate Turn-around' case, through the  `Gradual Turn-around' case, to the 
`Uniformly Accelerating' case. We show that this definition of simultaneity is independent of choice of coordinates, and assigns a {\bf unique} 
time to any event (with which the travelling twin can send and receive signals), resolving some common misconceptions.
\end{abstract}


\section*{INTRODUCTION}

It might seem reasonable to suppose that the twin paradox has long 
been understood, and that no confusion remains about it. Certainly there is 
no disagreement about the relative aging of the two twins, so 
there is no `paradox'. There is also no confusion over `when 
events are seen' by the two twins, or over the description by 
the stay-at-home twin (Alex say), of `when events happened'. These aspects of the twin paradox are treated in the standard texts\cte{Bohm,Dinverno,Leo,Pauli}. 

	However, 
the description of `when 
events happened' according to the travelling twin (Barbara say) seems 
never to have been fully  
settled\cte{Unruh2,Debs}. In textbook treatments, Barbara's hypersurfaces 
of simultaneity, which define `when events happened' according to her, have 
consistently been misrepresented or ignored. A common diagram for such hypersurfaces\cte{Bohm,Dinverno,Leo,Debs} 
is shown in Figure 1. 

\begin{figure}[h]
\vspace{-.3cm}
\center{\epsfig{figure=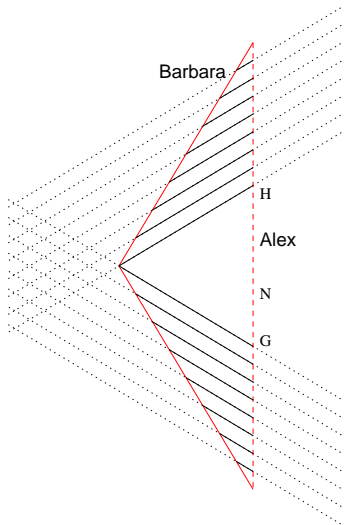, width=4.5cm}}
\caption{{\footnotesize A typical `textbook illustration' of the hypersurfaces of simultaneity of the travelling twin in the twin paradox.}}
\vspace{-.3cm}
\end{figure}

Bohm\cte{Bohm} and Sartori\cte{Leo} claim that points G and 
H appear to the travelling twin as simultaneous, while Bohm\cte{Bohm} goes on to say that 
``after the acceleration at E an event such as N is ascribed a smaller time coordinate than it had before''!  Other authors\cte{Leo,Dinverno}, noticing this problem, claim that its resolution lies in an analysis of the interval 
that the 
travelling twin must take to 
turn around (at a finite acceleration), and that during this period of acceleration Barbara's hypersurfaces of simultaneity `sweep 
round' from EG to FH, as shown in Figure 2.

\begin{figure}[h]
\vspace{-.3cm}
\center{\epsfig{figure=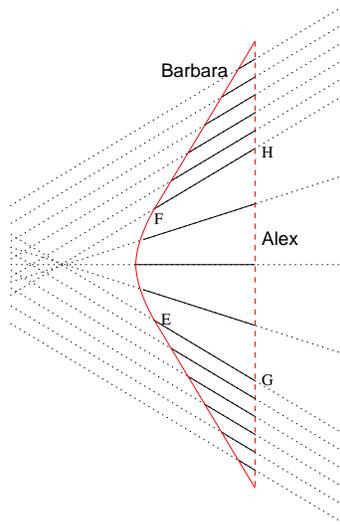, width=4.5cm}}
\caption{{\footnotesize Another typical `textbook illustration', in which Barbara's hypersurfaces of simultaneity `sweep round' from EG to FH during her period of acceleration.}}
\vspace{-.3cm}
\end{figure}

Although this period of acceleration can indeed fix the gap between G 
and H, it cannot resolve the more serious problem (mentioned also in Marder\cte{Marder} and 
in Misner et al.\cte{Misner}) which occurs to Barbara's left. Here her 
hypersurfaces of simultaneity are overlapping, and she assigns three times to every 
event! Also, if Barbara's hypersurfaces of simultaneity at a certain time depend so
sensitively on her instantaneous velocity as these diagrams suggest, then she 
would be forced to conclude that the distant planets 
swept backwards and forwards in time whenever she went dancing!

	The path through this confusion existed already in Einstein's original 
paper\cte{Ein}, and was popularised by Bondi in his work on `k-calculus'. It lies 
in the correct application of `radar time' (referred to as `M\"{a}rzke-Wheeler 
Coordinates' in Pauri et al.\cte{Pauri}). This concept is not new. Indeed 
Bohm\cte{Bohm} and D'Inverno\cte{Dinverno} both devote a whole 
chapter to k-calculus, and use `radar time' (not under that name) to derive the 
hypersurfaces of simultaneity of inertial observers. However, both authors then 
apply this definition wrongly to the travelling twin. 

	In the present paper we recall the definition of `radar time' (and  
related `radar distance') and emphasise that this definition applies not just 
to inertial observers, but to any observer in any spacetime. We then use radar time to 
derive the hypersurfaces of simultaneity for a class of travelling twins, from 
the `Immediate Turn-around' case, through the  `Gradual Turn-around' case, to the 
`Uniformly Accelerating' case. (The `Immediate Turn-around' and `Uniformly Accelerating' 
cases are also discussed in Pauri et al.\cte{Pauri}.) We show that in all cases this 
definition assigns a {\bf unique} 
time to any event with which Barbara can send and receive signals, and that this 
assignment is independent of any choice of coordinates. We then
demonstrate that brief periods of acceleration have negligible effect on 
the radar time assigned to distant events, in contrast with the sensitive dependence of the hypersurfaces implied by Figures 1 and 2. By viewing the situation in different coordinates 
we further demonstrate the coordinate independence of radar time, and note that there is no observational difference between the interpretations in which the differential aging is `due to Barbara's 
acceleration' or `due to the gravitational field that Barbara sees because 
of this acceleration'.

\section*{RADAR TIME AND RADAR DISTANCE}

Consider an observer travelling on path $\gamma: x^{\mu} = x^{\mu}_{(\gamma)}(\tau)$ with proper time $\tau$. Define:
\begin{align}
\tau^{+}(x) & \equiv \mbox{ (earliest possible) proper time at which a light ray} \notag \\
& \mbox{(technically, a null geodesic) leaving point $x$ could intercept $\gamma$. } \notag \\
\tau^{-}(x) & \equiv \mbox{ (latest possible) proper time at which a light ray} \notag \\
&  \mbox{(null geodesic) could leave $\gamma$, and still reach point $x$. } \notag \\
\tau(x) & \equiv \half (\tau^{+}(x) + \tau^{-}(x)) \hs{1} = \mbox{ `radar time'.} \notag \\
\rho(x) & \equiv \half (\tau^{+}(x) - \tau^{-}(x)) \hs{1} = \mbox{ `radar distance'.} \notag \\
\Sigma_{\tau_0} & \equiv \{x: \tau(x) = \tau_0 \} \notag \\
 & \hs{-1} = \mbox{ observer's `hypersurface of simultaneity at time $\tau_0$'. } \notag \end{align}

\begin{figure}[h]
\vspace{-.3cm}
\center{\epsfig{figure=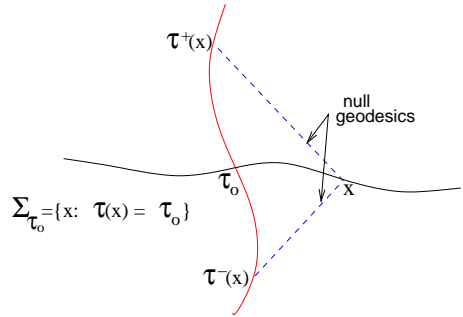, width=6cm}}
\caption{{\footnotesize Schematic of the definition of `radar time' $\tau(x)$.}}
\vspace{-.3cm}
\end{figure}

This definition was made popular by Bondi in his work on 
special relativity and k-calculus\cte{Bondi,Bohm,Dinverno}, and is 
essentially a rearrangement of the formula $t - t^- = t^+ - t$ from Einstein's 
original paper\cte{Ein} (written as $t_B - t_A = t'_A - t_B$ in Einstein's 
notation). This formula simply says that an observer can assign a time to a distant event by 
sending a light signal to the event and back, and averaging 
the (proper) times of sending and receiving. It is clearly applicable to any 
observer in any gravitational background. However, the idea of defining 
hypersurfaces of simultaneity in 
terms of `radar time' has seldom been applied to non-inertial observers or to non-flat spacetimes. This 
is perhaps due to Bondi's claim that\cte{Bondi} ``how a clock reacts to acceleration 
 is utterly dependent on how the clock is 
constructed''. This claim has since been refuted by experiment (see \cte{Marder}, or \cte{Dav3} 
pp 144,145, and references therein); moreover the assumption that `suitable clocks' 
will behave identically under acceleration (or under gravitational fields) is a 
basic premise of general relativity, without which proper time would  
have no physical meaning. There is therefore no reason not to apply `radar 
time' to accelerating observers or to curved spacetimes. 

Since radar time can be defined without any mention of 
coordinates it is, by construction, independent of our choice 
of coordinates. It is single-valued, agrees with proper time on the observers path, and is 
invariant under `time-reversal' - that is, under reversal of the 
sign of the observer's proper time.

\section*{THE TRAVELLING TWIN}

Consider the standard `Immediate Turn-around' case, for which Barbara's path is given by
\begin{equation} (t_{B}(\tau),x_{B}(\tau)) = \begin{cases} 
(\cosh(\alpha) \tau \ , \  \sinh(\alpha) \tau) & (\tau > 0)  \\
(\cosh(\alpha) \tau \ , \  -\sinh(\alpha) \tau) &  (\tau < 0)  \\ \end{cases} \notag \end{equation}
or:
\begin{equation} (u_{B}(\tau), v_{B}(\tau)) = \begin{cases} 
(e^{\alpha} \tau \ , \  - e^{-\alpha} \tau) & (\tau > 0)  \\ 
(e^{-\alpha} \tau \ , \ - e^{\alpha} \tau) & (\tau < 0)  \\  \end{cases} \notag \end{equation}
where $s = \ta$ is Barbara's speed (in units with $c=1$), $\alpha$ is her `rapidity',  and we have defined coordinates $(u,v)$ by $u \equiv x + t$ and $v \equiv x - t$. This situation is shown in Figure 4 below.

\begin{figure}[h]
\center{\epsfig{figure=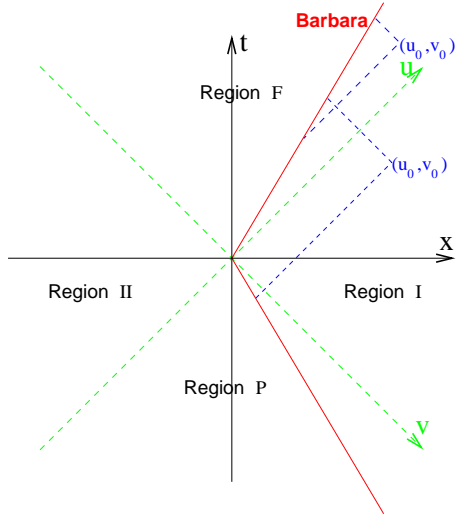, width=6cm}}
\caption{{\footnotesize The travelling twin in the `Immediate Turn-around' case.}}
\vs{-.1}
\end{figure}

To deduce the radar time of a point $(u_0,v_0)$, it is 
convenient to consider 4 different regions, 
depending on the signs of $u_0$ and $v_0$.

\subsubsection*{Region I: $u_0 > 0$ and $v_0 > 0$}

	Clearly, $\tau^+(u_0,v_0)$ is determined by putting $u_0 = u_B(\tau^+) = e^{\alpha} \tau^+$ (where we have used $u_B(\tau^+) = e^{\alpha} \tau^+$ since we 
 see from Figure 4 that $\tau^+ > 0$). Similarly, $\tau^-(u_0,v_0)$ is determined by putting $v_0 = v_B(\tau^-) = - e^{\alpha} \tau^-$ (since $\tau^- < 0$). Hence:
\begin{align} \tau(u_0,v_0) & = \half (e^{-\alpha} u_0 - e^{-\alpha} v_0 ) \notag \\
& = e^{-\alpha} t_0 \end{align}
	in this region.

\subsubsection*{Region F: $u_0 > 0$ and $v_0 < 0$}

	Consider first the case where $(u_0,v_0)$ is to Barbara's right, as shown 
in Figure 4. The calculation of $\tau(u_0,v_0)$ proceeds as in the previous case, 
the only difference being that, since $\tau_- > 0$, we now use $v_B(\tau^-) = 
- e^{-\alpha} \tau^-$ (the exponent has changed sign). Hence we have
\begin{align} \tau(u_0,v_0) & = \half (e^{-\alpha} u_0 - e^{\alpha} v_0 ) \notag \\
& = \ca t_0 - \sa x_0 \label{eq:tauRegF}\end{align}

	If $(u_0,v_0)$ is to Barbara's left (but still in region F) then the roles played by $\tau^+$ and $\tau^-$ are reversed, but since the definition of $\tau$ is symmetric under this change it follows that (\ref{eq:tauRegF}) remains unchanged, and holds throughout region F.

	Regions II and P are treated similarly, and we find:

\begin{equation} \tau(t,x) = \begin{cases} 
t e^{-\alpha} & \mbox{ region I} \\
t e^{\alpha} & \mbox{ region II} \\
\ca t - \sa x & \mbox{ region F} \\
\ca t + \sa x & \mbox{ region P} \\ \end{cases} \notag \end{equation}

The `hypersurface of simultaneity at time $\tau_0$', denoted $t_{\tau_0}(x)$ is now given by putting $\tau(t,x)=\tau_0$ and rearranging. This gives:
\begin{equation} t_{\tau_0}(x) = \begin{cases} 
\tau_0 e^{\alpha} & \mbox{ region I} \\
\tau_0 e^{- \alpha} & \mbox{ region II} \\
\seca \tau_0 + \ta x  & \mbox{ region F} \\
\seca \tau_0 - \ta x & \mbox{ region P} \\ \end{cases} \notag \end{equation}
as shown for various $\tau_0$ in Figure 5.

\begin{figure}[h]
\center{\epsfig{figure=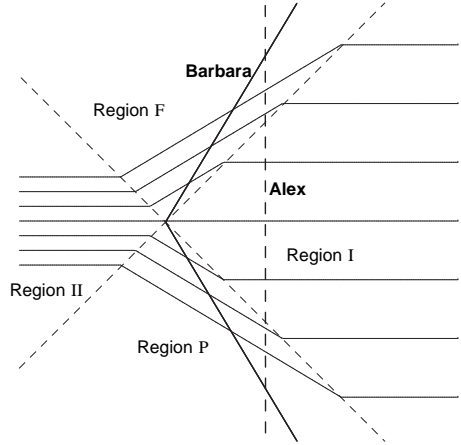, width=6cm}}
\caption{{\footnotesize Hypersurfaces of simultaneity of the travelling twin in the `Immediate Turn-around' case.}}
\vs{-.1}
\end{figure}

We now see that in region F, where Barbara is indistinguishable from an 
inertial observer with velocity $s$, her hypersurfaces of simultaneity are 
exactly those appropriate to such an inertial observer; while in region P, where she 
is indistinguishable from an inertial observer with velocity $-s$, her 
hypersurfaces are those appropriate to that inertial observer. This is not  
true in 
regions I and II however, for which she has  
differing velocities when she sends her signal (at time $\tau^-$) when she receives it back (at time $\tau^+$). In these regions her 
hypersurfaces of simultaneity are flat (as expected for symmetry 
under time reversal). Hence she assigns a {\bf unique} time to 
each event, avoiding the problems that appear in the common `textbook 
illustrations' (Fig. 1, 2). 

	We now consider the effect on these conclusions of 
a small period of acceleration at the point of turn-around.

\section*{GRADUAL TURN-AROUND}

We allow Barbara to turn around at finite acceleration $a$, for a proper time $\tau_c = \frac{\alpha}{a}$. Barbara's trajectory is now given by:

\begin{equation} t_{B}(\tau) = \begin{cases} 
\ct (\tau - \tau_c) + \frac{1}{a} \st & (\tau \geq \tau_c) \\
\frac{1}{a} \sinh(a \tau) & (|\tau| \leq \tau_c) \\ 
\ct (\tau + \tau_c) - \frac{1}{a} \st & (\tau \leq -\tau_c) \\ \end{cases} \notag \end{equation}

\begin{equation} x_{B}(\tau) = \begin{cases} 
 \st (\tau - \tau_c) + \frac{1}{a} \ct & (\tau \geq \tau_c) \\
\frac{1}{a} \cosh(a \tau) & (|\tau| \leq \tau_c) \\ 
 - \st (\tau + \tau_c) + \frac{1}{a} \ct & (\tau \leq -\tau_c) \\ \end{cases} \notag \end{equation}
and her final speed is $s = \tanh(a \tau_c)$. In null coordinates this is:

\begin{align} (u_{B}(\tau),v_B(\tau)) & \notag \\
&  \hs{-1.5} =  \begin{cases} \hs{-.1} 
( e^{a \tau_c} (\tau \hs{-.1} - \hs{-.1} \tau_c \hs{-.1} + \hs{-.1} 
\frac{1}{a}) \ , \ - e^{-a \tau_c} (\tau \hs{-.1} - \hs{-.1} \tau_c 
\hs{-.1} - \hs{-.1} \frac{1}{a}) ) &  (\tau \geq \tau_c) \\
(\frac{1}{a} e^{a \tau} \ , \ \frac{1}{a} e^{- a \tau}) & (|\tau| \leq \tau_c) \\ 
(e^{-a \tau_c} (\tau \hs{-.1} + \hs{-.1} \tau_c \hs{-.1} + \hs{-.1} 
\frac{1}{a}) \ , \ - e^{a \tau_c} (\tau \hs{-.1} + \hs{-.1} \tau_c 
\hs{-.1} - \hs{-.1} \frac{1}{a}) ) & (\tau \leq -\tau_c) \\ \end{cases} 
\notag \end{align}

We now have nine regions to consider, separated by $u = e^{\pm a \tau_c}$ and $v = e^{\pm a \tau_c}$. The calculation is straightforward, and yields hypersurfaces of simultaneity of the form:

\begin{equation} t_{\tau_0}(x) = \begin{cases} 
\tau_0 e^{\pm a \tau_c} & \mbox{ regions I, II} \\
\sect (\tau_0 \pm \tau_c) \mp \tant x & \mbox{ regions P, F} \\
x \tanh(a \tau_0) & \mbox{ region U} \\ 
\end{cases} \notag \end{equation}
while in the other 4 regions these hypersurfaces are given implicitly by:
\begin{align}
\log (a|x & + t_{\tau_0}(x)|) + a (t_{\tau_0}(x) - x) e^{\pm a \tau_c} \notag \\
&  = 2 a \tau_0 \mp a \tau_c - 1 \hs{1} \mbox{ regions FII, PI} \notag \\
\log (a|x & - t_{\tau_0}(x)|) - a (t_{\tau_0}(x) + x) e^{\pm a \tau_c} \notag \\ 
& = - 2 a \tau_0 \mp a \tau_c - 1 \hs{.7} \mbox{ regions FI, PII} \notag 
\end{align}

\begin{figure}[h]
\center{\epsfig{figure=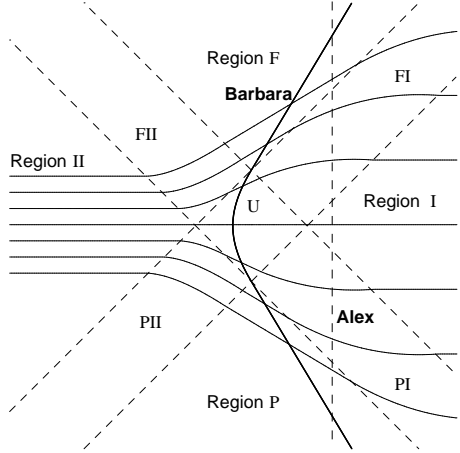, width=6cm}}
\caption{{\footnotesize Hypersurfaces of simultaneity of the travelling twin in the `Gradual Turn-around' case.}}
\vs{-.1}
\end{figure}

	These hypersurfaces are shown in Figure 6. As hoped, there is 
very little difference (for small $\tau_c$) between the `Immediate Turn-around' and 
`Gradual Turn-around' cases. The hypersurfaces in regions I and II are identical to 
the previous case, while in regions  F and P the only difference is that a slightly 
different value of $\tau_0$ is assigned to each hypersurface, due to the slight 
difference in turn-around time. This contrasts with the situation described in 
many textbooks\cte{Marder,Dinverno} and depicted in Figure 2. 

	It is interesting to 
consider the limit $\tau_c \rightarrow \infty$, corresponding to a uniformly 
accelerating observer. In this case we have simply:

\begin{align} (t_{B}(\tau),x_B(\tau)) & = \left( \frac{1}{a} \sinh(a \tau) \ , \ \frac{1}{a} \cosh(a \tau) \right) \notag \\
\mbox{ or } (u_{B}(\tau),v_B(\tau)) & = \left( \frac{1}{a} e^{a \tau} \ , \ \frac{1}{a} e^{-a \tau} \right) \notag \end{align}

A simple calculation yields radar time 
 and radar distance:

\begin{align} \tau(t,x) & = \frac{1}{2a} \log\left(\frac{x + t}{x - t}\right) \notag \\ 
\rho(t,x) & = \frac{1}{2a} \log(a^2 (x^2 - t^2)) \notag \end{align}

The hypersurfaces of simultaneity are therefore given by
$$ t_{\tau_0}(x) = x \tanh(a \tau_0) $$
	just as in region $U$ of Fig. 6.

\begin{figure}[h]
\center{\epsfig{figure=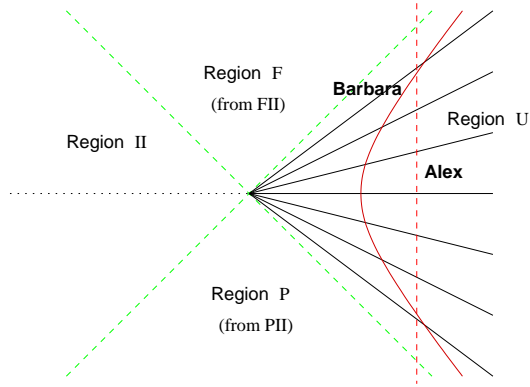, width=7cm}}
\caption{{\footnotesize Hypersurfaces of simultaneity of a uniformly accelerating observer.}}
\end{figure}

	$\rho$ and $\tau$ are simply Rindler coordinates now, and 
cover only region U. This is as expected, since Barbara cannot send a signal to any events in regions P or II, while no signals could reach Barbara from regions F or II.

\section*{GRAVITY DOESN'T MATTER}

	It is often said of the twin paradox that\cte{Pauli} ``a complete explanation of the problem 
can only be given within the 
framework of general relativity''. However, as 
we have just shown, Barbara's hypersurfaces of simultaneity depend only on the kinematics involved, and can be fully understood 
without resorting to general relativity. To understand why `equivalent 
gravitational fields' are some times invoked in the literature, consider the 
`Uniformly Accelerating' case, as described in $(\tau,\rho)$ coordinates. This 
is shown in Figure 8. 

\begin{figure}[h]
\center{\epsfig{figure=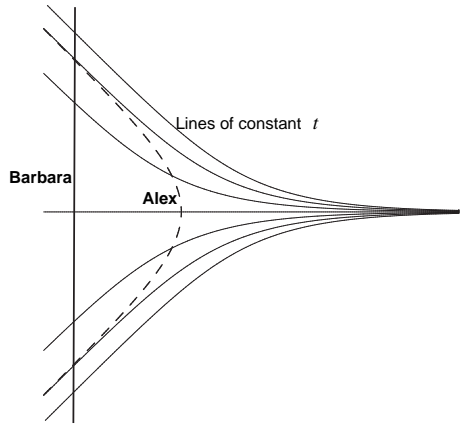, width=6cm}}
\caption{{\footnotesize Alex's hypersurfaces of simultaneity as described in Barbara's $(\tau, \rho)$ coordinates, for the `Uniformly Accelerating' case.}}
\end{figure}

The metric in these coordinates is $ d s^2 = e^{2 a \rho} (d \tau^2 - d \rho^2) $. 
Although Barbara has a `straight trajectory' in these coordinates, she is still 
undergoing an acceleration $a$, just as we undergo a constant acceleration g 
through being `held up' against the earth's gravitational field. (Of course 
Barbara is here being `pulled down' rather than pulled `towards the earth's centre', 
so no tidal forces are involved.) Although Alex's trajectory appears curved in these 
coordinates, he is still inertial, since his trajectory is geodesic in this 
metric. Alex and Barbara's proper times are unaltered by this change of coordinates, 
so that it is still Alex who ages more during his absence from Barbara. Some of 
the lines of constant $t$, which represent Alex's hypersurfaces of simultaneity, fail 
to intercept his trajectory in Figure 8. This is because Barbara's 
$(\tau, \rho)$ coordinates cover only region U of Figure 7, so any line of 
constant $t$ which intercepts Alex's trajectory in region F or P will not 
have the interception visible in Figure 8.

	A similar coordinate transformation can be performed for the `Immediate Turn-around' case. In $(\tau,\rho)$ coordinates this is shown in Figure 9.

\begin{figure}[h]
\center{\epsfig{figure=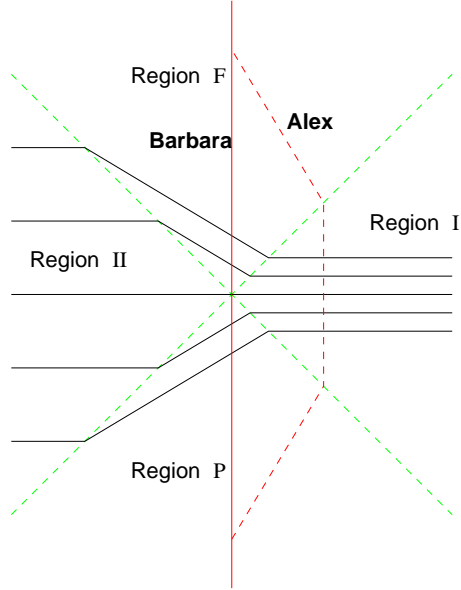, width=6cm}}
\caption{{\footnotesize Alex's hypersurfaces of simultaneity, as described in Barbara's $(\tau, \rho)$ coordinates, for the `Immediate Turn-around' case.}}
\end{figure}

The metric in these coordinates is:

\begin{equation} d s^2 = \begin{cases} 
 e^{-2 \alpha} (d \tau^2 - d \rho^2) & \mbox{region II} \\
 (d \tau^2 - d \rho^2) & \mbox{regions F,P} \\
 e^{2 \alpha} (d \tau^2 - d \rho^2) & \mbox{region I} \\
 \end{cases} \notag \end{equation}
which is Minkowski in all 4 regions, but with a shock-like scale discontinuity 
along the lines $x = \pm t$ which causes geodesics to `turn left' upon crossing 
these lines. In these coordinates Alex's hypersurfaces of simultaneity look much 
as Barbara's did in the original coordinate system. It is still Alex who is inertial, 
however. Graphing the results in different coordinate systems makes no difference to 
this conclusion. Also, since the definition of radar time is independent of the choice 
of coordinates, then any statement that Barbara makes of the form `Alex did this at 
this time' (according to her), or any statement that Alex makes of the form `Barbara did 
this at this time' (according to him), will be unchanged by changes in coordinates.

\section*{CONCLUSION}

	We have demonstrated that `radar time' is applicable to arbitrarily 
moving observers, and we have used it to find the hypersurfaces of simultaneity of the 
travelling twin in the `twin paradox'. Radar time allows the travelling twin to assign times to distant events in a fully coordinate invariant fashion. We have considered a general class 
of twins, varying from the `Immediate Turn-around' case to the `Uniformly 
Accelerating' observer, and have highlighted some common misconceptions. 

	Throughout this paper we have made a convenient but unrealistic 
simplification. It is clear from the trajectories shown in the previous 
figures that Barbara and Alex could not possibly be twins, since they never 
spend more than an instant in each other's company. More realistically, Barbara's 
trajectory should coincide with Alex's until she begins her journey, and 
should coincide again after her return. The hypersurfaces of simultaneity for this more realistic trajectory are shown in Figure 10.

\begin{figure}[h]
\center{\epsfig{figure=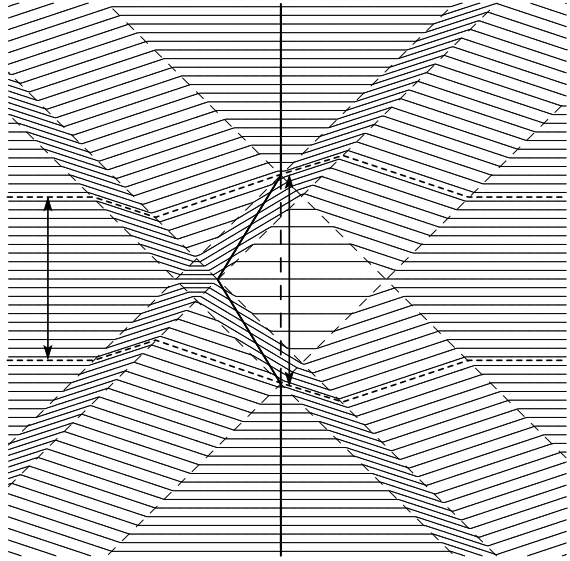, width=7.5cm}}
\caption{{\footnotesize Hypersurfaces of simultaneity of the travelling twin in the more realistic case, in which their trajectories coincide before and after the journey.}}
\end{figure}

In the {\it causal envelope}\cte{Pauri} of Barbara's journey (that is: the intersection of the causal future of her departure and the causal past of her return, being the middle four regions in Figure 10) these hypersurfaces are identical to those of Figure 5. Readers might enjoy interpreting the rest of this diagram, convincing themselves that the length of the left arrow represents Barbara's travelling time and that the length of the middle arrow represents Alex's, and verifying that the ratio of these is indeed $\sqrt{1 - s^2}$.

\section*{ACKNOWLEDGEMENTS}

We thank Dr Anton Garrett for his many helpful comments on this manuscript. Carl Dolby thanks Dr Henry Levy for 
his enthusiastic teaching of k-calculus, and thanks Cavendish 
Astrophysics, and Trinity College, Cambridge, for financial support.

\end{document}